\documentclass[aip,reprint,jap]{revtex4-1}

\newcommand{\be}{\begin{equation}}
\newcommand{\ee}{\end{equation}}

\newcommand{\dV}{\,\text{d}V}

\usepackage[utf8]{inputenc}
\usepackage[T1]{fontenc}
\usepackage{bm}
\usepackage{amsmath}   
\usepackage{amsfonts}
\usepackage{siunitx}    
\usepackage{graphicx} 

\begin{document}

\title{Topological characterization of Hopfions in finite-element micromagnetics}

\author{Louis Gallard}
\author{Riccardo Hertel}
 \email{riccardo.hertel@ipcms.unistra.fr}
\affiliation{ Université de Strasbourg, CNRS, Institut de Physique et Chimie des Matériaux de Strasbourg, F-67000 Strasbourg, France
}

\begin{abstract}
Topological magnetic structures, such as Hopfions, are central to three-dimensional magnetism, but their characterization in complex geometries remains challenging. We introduce a robust finite-element method for calculating the Hopf index in micromagnetic simulations of three-dimensional nanostructures. By employing the Biot-Savart form for the vector potential, our approach ensures gauge-invariant results, even in multiply connected geometries like tori. A novel variance-based correction scheme significantly reduces numerical errors in highly inhomogeneous textures, achieving accurate Hopf index values with fast mesh-dependent convergence. We validate the method using an analytically defined Hopfion structure and demonstrate its ability to detect topological transitions through a simulation of a Hopfion’s field-induced destruction into a toron, marked by an abrupt change in the Hopf index.
This method enables precise quantification of topological features in complex three-dimensional magnetic textures forming in finite-element micromagnetic simulations, offering a powerful tool for advancing topological magnetism studies in general geometries.
\end{abstract}


\maketitle 

\section{Introduction}

Over the past years, research in magnetism has seen remarkable progress in three-dimensional (3D) nanomagnetism \cite{fernandez-pacheco_three-dimensional_2017} in terms of nanofabrication, imaging techniques \cite{donnelly_three-dimensional_2017}, and simulation capabilities. The intense scientific activity in this domain marks a clear departure from previous extensive studies in magnetism centered on investigating thin and ultrathin magnetic films \cite{bland_ultrathin_1995}. In addition to sparking growing interest in the magnetic properties of nanostructures featuring complex 3D geometries \cite{cheenikundil_switchable_2021,golebiewski_collective_2024}, the recently gained access to the third dimension \cite{fischer_launching_2020} has fueled research on characteristic 3D magnetic textures absent in thin-film geometries, like skyrmion tubes \cite{birch_real-space_2020}, Bloch points \cite{milde_unwinding_2013} (including variants thereof---such as torons\cite{leonov_homogeneous_2018} or chiral bobbers \cite{zheng_experimental_2018}), or Hopfions \cite{kent_creation_2021,rybakov_magnetic_2022,zheng_Hopfion_2023}. The latter are remarkably complex textures exhibiting knots in the magnetization vector field. Hopfions have recently received considerable attention due to their intriguing properties and their possible use in future spintronic devices\cite{gobel_topological_2020}. 

Micromagnetic simulations play a crucial role in developing a thorough understanding of these topological textures, which are vastly underexplored compared to skyrmions that can be considered to be their two-dimensional (2D) counterparts. Specifically, finite-element micromagnetic simulations are essential in numerical studies, since the method allows for accurate consideration of details of the 3D sample geometry, including curved and oblique surfaces, which are inherently problematic in the case of finite-difference methods\cite{garcia-cervera_accurate_2003}. Hopfions, like skyrmions, are topologically protected magnetization structures of nanometric size with particle-like properties. In contrast to 2D topological magnetic textures, such as vortices or skyrmions, which are easily identifiable in magnetic configurations, Hopfions are challenging to detect due to their complex 3D knotted vector field distribution embedded within the magnetic volume. Even in micromagnetic simulations, where magnetic structure data is readily accessible and vector field visualization is straightforward, unambiguously identifying Hopfions remains difficult.

A reliable means to quantitatively detect Hopfions in micromagnetically simulated 3D textures consists in determining the Hopf index, a topological invariant
\cite{whitehead_expression_1947,faddeev_stable_1997,wilczek_linking_1983,nicole_solitons_1978}. Numerically calculating this invariant can entail various complications and pitfalls regarding accuracy, efficiency, and gauge invariance. In this article, we present the implementation of a flexible and robust finite-element algorithm to calculate the Hopf index. Particular attention is paid to the question of gauge invariance of the computed values. We critically discuss aspects of numerical accuracy when calculating the Hopf index and introduce an efficient method to minimize discretization errors. Finally, we present an application example of the method to monitor
magnetic Hopfions in simulations.

\section{Hopf invariant\label{hopf_ind_sec}}

The Hopf index is a topological invariant that belongs to the homotopy group $\pi_{3}(\mathbb{S}^{2})$. It has a non-zero integer value for Hopfions and can be calculated through Whitehead’s formula \cite{whitehead_expression_1947}.

\be\label{hopf}
N = -\int\bm{J}\cdot\bm{A}\dV \quad,
\ee
where $\bm{A}$ is a vector potential satisfying

\be\label{curlA}
\nabla\times\bm{A}=\bm{J} \quad.
\ee

The vector $\bm{J}$ is defined through the normalized magnetization $\bm{m}=\bm{M}/|M|$ according to

\be\label{topo_curr}
J_i= \frac{1}{8\pi}\epsilon_{ijk} \bm{m}\cdot\left(\frac{\partial\bm{m}}{\partial x_j}\times\frac{\partial\bm{m}} {\partial x_k}\right)
\ee

Various terminologies are used throughout the literature to refer to the vector field $\bm{J}$ or variants thereof, including {\em emergent field} \cite{liu_binding_2018,tai_static_2018}, {\em vorticity} \cite{komineas_topology_1996}, or {\em gyrovector} \cite{thiele_steady-state_1973,rybakov_magnetic_2022}. In the present study, we adopt the term {\em topological current}\cite{wilczek_linking_1983,han_skyrmions_2017} for $\bm{J}$. 
Importantly, the topological current $\bm{J}$ is divergence-free\footnote{The property $\nabla\cdot\bm{J}=0$ is central to this formalism, as it allows to introduce a vector potential according to eq.~(\ref{curlA}). This condition of a solenoidal field $\bm{J}$ is not fulfilled in the case of Bloch points, where the magnetization's spatial gradients are undefined.} for any smooth magnetic texture of the unit vector field $\bm{m}$ \cite{komineas_topology_1996,han_skyrmions_2017}.

In addition to the integral definition of eq.~(\ref{hopf}), the Hopf index has a topological interpretation\cite{arnold_topological_2021} as the linking number of the preimages (regions within the domain in which the field has the same orientation) of two arbitrary points of $S^2$, i.e., the sphere representing the vector field's parameter space\cite{ackerman_diversity_2017}.
Hopfions have closed-loop preimages that are characteristically linked to each other, resulting in knotted field lines. 
The formalism related to the Hopf invariant is broadly applied to categorize knotted textures of vector fields in liquid crystals \cite{smalyukh_review_2020} and ferroic materials \cite{sutcliffe_skyrmion_2017,tai_static_2018,lukyanchuk_Hopfions_2020}.
More generally, the analysis of knotted fields finds application in a remarkably large variety of disciplines~\cite {ricca_knotted_2024}, including plasma physics \cite{moffatt_degree_1969}, hydrodynamics \cite{moreau_constantes_1961}, photonics
\cite{dennis_isolated_2010,barnett_lines_2023}, and quantum field theory\cite{rajaraman_solitons_1982,arafune_topology_1975}. 

To represent a physically meaningful quantity, the Hopf index (\ref{hopf}) must be gauge invariant, i.e., remain unchanged under a transformation $\bm{A}\rightarrow\bm{A}+\bm{\nabla}\chi$. Given the explicit appearance of the vector potential $\bm{A}$ in the integrand of equation (\ref{hopf}), the invariance of the Hopf index is not immediately apparent. Indeed, the topological charge density $\rho_H=\bm{J}\cdot\bm{A}$ is not uniquely determined due to its lack of gauge invariance\cite{nicole_solitons_1978}. Nevertheless, in many cases, the Hopf index according to eq.~(\ref{hopf}) {\em is} gauge invariant. The necessary conditions for this are that the domain $\Omega$ over which the integral extends is singly-connected \cite{berger_topological_1984,woltjer_theorem_1958,moreau_constantes_1961} and that the topological current $\bm{J}$ is either zero or tangential to the integration region's surface $\partial\Omega$, i.e. $\bm{n}\cdot\bm{J}=0$, where $\bm{n}$ is the surface normal vector\cite{berger_topological_1984}.

The case of multiply connected geometries, such as a torus, has been discussed in the context of helicity in magnetohydrodynamics \cite{mactaggart_magnetic_2019, bevir_relaxation_1980, ricca_knotted_2024}.
These studies show that additional line integrals along closed paths on the surface  $\partial\Omega$  are required to render the Hopf index gauge invariant. The ``Biot-Savart helicity'' plays an important role in this context, where the vector potential $\bm{A}$ is calculated through the Biot-Savart operator:
\be\label{biotsavart}
    \bm{A}(\bm{r})=\nabla \times \frac{1}{4\pi}\int\frac{\bm{J}(\bm{r'})}{\lvert \bm{r} -\bm{r'}\lvert}{\rm d}V'    
\ee
When the Hopf index, as defined in eq.~(\ref{hopf}), is calculated using a vector potential $\bm{A}$ derived in this manner, it is equivalent to the gauge-invariant form. This eliminates the need for the additional line integrals in multiply connected geometries. In this formalism, the gauge is chosen such that the potential satisfies
$\bm{\nabla}\cdot\bm{A}=0$ and $\left|\bm{A}\right|$ vanishes at infinity. Here, the vector potential $\bm{A}$ is linked to the topological current $\bm{J}$ in a manner analogous to a magnetic field generated by an electric current density distribution $\bm{j}$, thus establishing a formal analogy between topological and electrical currents. 

\section{Method\label{method_sect}}
We use a custom-developed finite-element method (FEM) software to numerically compute the Hopf index in 3D ferromagnetic nanostructures of arbitrary shape. The domain $\Omega$, encompassing the entire volume $V$ of the nanostructure, is discretized into tetrahedral simplex elements. The magnetization vector field is defined at the nodes and interpolated within each element to obtain a piecewise linear representation.

To calculate the Hopf index, we first compute the topological current as defined in Eq.~(\ref{topo_curr}). Within each element, first-order spatial derivatives of the discretized magnetization vector field are calculated using linear shape functions, yielding piecewise constant derivative values. At the nodes, the derivative is approximated by a volume-weighted average of the derivatives from all adjacent elements. This approach corresponds to a mass-lumped representation of the weak form for the first derivative.

For the calculation of the vector potential $\bm{A}$, we adopt the Biot-Savart form described in Eq.~(\ref{biotsavart}). However, direct integration of this equation is computationally inefficient due to the twofold volume integral required across the domain $\Omega$. Instead, we solve an equivalent set of partial differential equations derived by taking the curl of Eq.~(\ref{curlA}) and enforcing the Coulomb gauge $\bm{\nabla}\cdot\bm{A}=0$, yielding the Poisson equation:
\cite{wilczek_linking_1983}
\be
    \Delta \bm{A} = -\nabla \times \bm{J}
    \label{Poisson_A}
\ee
In a component-wise representation, the three differential equations can be written as
\be\label{Poisson_a_comp}
\Delta A_i = \nabla\cdot\left(\bm{e}_i\times\bm{J}\right) , \quad i=x,y,z
\ee
where $\bm{e}_i$ is the Cartesian unit vector along the $i$-direction.
At the domain boundary $\partial\Omega$, the following jump condition applies:
\be\label{jumpA}
(\bm{n}\cdot\nabla)\bm{A}_\text{in}-(\bm{n}\cdot\nabla)\bm{A}_\text{out}=\bm{J}\times\bm{n}
\ee 
where $\bm{n}$ is the outward oriented surface normal vector and $\bm{A}_\text{in}$, $\bm{A}_\text{out}$ are the inward and outward limit values of the vector potential at the surface $\partial\Omega$.
Equations (\ref{Poisson_A}) and (\ref{jumpA}), combined with the Coulomb gauge $\bm{\nabla}\cdot\bm{A}=0$ and the boundary condition $\lim_{\left|\bm{x}\right|\to\infty}\bm{A}(\bm{x})=\bm{0}$, make the problem formally analogous to the Oersted field calculation for a current-carrying conductor. We can, in fact, use precisely the methods described in Ref.~\cite{hertel_hybrid_2014} by replacing $\bm{H}\rightarrow\bm{A}$ and $\bm{j}\rightarrow\bm{J}$. Specifically, by applying these methods, we solve the open-boundary problem of Eq.~(\ref{curlA}) with a hybrid finite-element/boundary-element (FEM/BEM) algorithm implementing the methodology described by Fredkin and Koehler\cite{fredkin_hybrid_1990}. The boundary integral of the BEM part is numerically computed using $\mathcal{H}2$-type hierarchical matrices, as described in Ref.~\cite{hertel_hybrid_2014}. The Poisson equations (\ref{Poisson_a_comp}) are solved using an algebraic multigrid method with GPU (graphical processing unit) acceleration \cite{demidov_amgcl_2019}. 
Once the vector potential $\bm{A}$ and the topological current $\bm{J}$ are computed at each discretization point, the numerical integration over the domain $\Omega$, as described on the right-hand side of Eq.~(\ref{hopf}), is straightforward.

\section{Analytic form of a Hopfion texture}
To validate our method and demonstrate its efficiency, we apply it to an idealized Hopfion structure, defined analytically by a set of equations and parameters. 
The following ansatz defines a Hopfion in an $xy$-torus of major radius $r_\text{maj}$ and minor radius $r_\text{min}$. 

\begin{align}
\label{bimeron}
        m_x &= \sin\left(\pi \frac{\rho}{r_\text{min}}\right) \cos\left(v\theta -u\phi + \alpha\right) \\
    m_y &= \sin\left(\pi\frac{\rho}{r_\text{min}}\right) \sin\left(v\theta - u\phi + \alpha\right) \\
    m_z &= m_0 \cos\left(\pi \frac{\rho}{r_\text{min}}\right)\label{bimeron_end}
\end{align}

with 
\begin{align}
    \rho &= \sqrt{z^2 + (\sqrt{x^2 + y^2} - r_\text{maj})^2} \\ \theta &= \arctan({y}/{x})\\ 
    \phi &= \arctan\left(\frac{z}{\sqrt{x^{2} + y^{2}} - r_\text{maj}}\right)
\end{align}

\begin{figure}[!ht]
    \includegraphics[width=\linewidth]{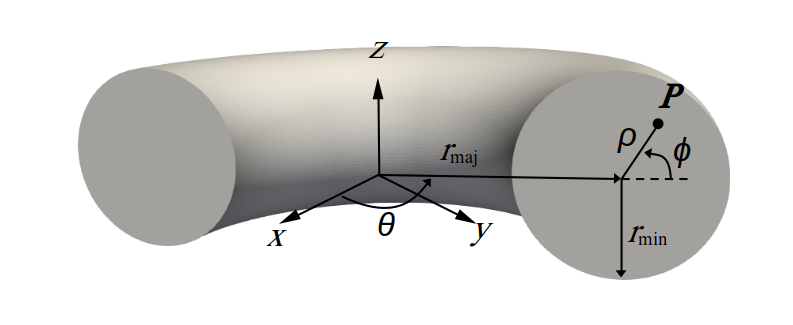}
    \caption{Schematic representation of the parameters and variables used in the analytic  Hopfion configuration, showing how the position of a point $P$ in the torus is defined through the coordinates $\theta$, $\rho$, and $\phi$.}
    \label{torus_labels}
\end{figure}

Here, $m_{0}$ defines the magnetization at the Hopfion's core ($\rho = 0$), where  $m_z(\rho=0) = m_0$, and at the torus surface ($\rho = r_\text{min}$), where $m_z(\rho=r_\text{min}) = -m_0$. Fig.~\ref{torus_labels} provides a schematic representation of the variables used in Eqs.~(\ref{bimeron})-(\ref{bimeron_end}).

Notably, while the inhomogeneous region of the magnetization is confined within the volume of the torus defined by $r_\text{maj}$ and $r_\text{min}$, the torus surface does not necessarily represent the physical boundary of the magnetic nanostructure. 
The magnetic texture described by this ansatz can be embedded into a homogeneous background magnetization in the $z$ direction with $m_z=-m_0$. 
In this case, the topological charge has compact support, meaning the topological features and the topological current $\bm{J}$ are confined within the torus and vanish outside.

In Eqs.~(\ref{bimeron})-(\ref{bimeron_end}), the parameter $\alpha$ introduces an additional phase for the in-plane magnetization components, allowing variation in the bimeron's orientation across cross-sections. The parameters $u$ and $v$ are integer numbers determining the Hopfion's chirality and order, with their values directly affecting the degree of winding, which is correlated with the Hopf index. Their impact on the vector field topology will be discussed below, in section ~\ref{ho_hopf_sec}. Setting $u=v=1$, the magnetic configuration is axially symmetric, with each poloidal cross-section (along the meridional plane) exhibiting a bimeron structure, i.e., a vortex-antivortex pair with opposite core polarizations\cite{gobel_topological_2020}. 

\begin{figure}[!ht]
\includegraphics[width=1\linewidth]{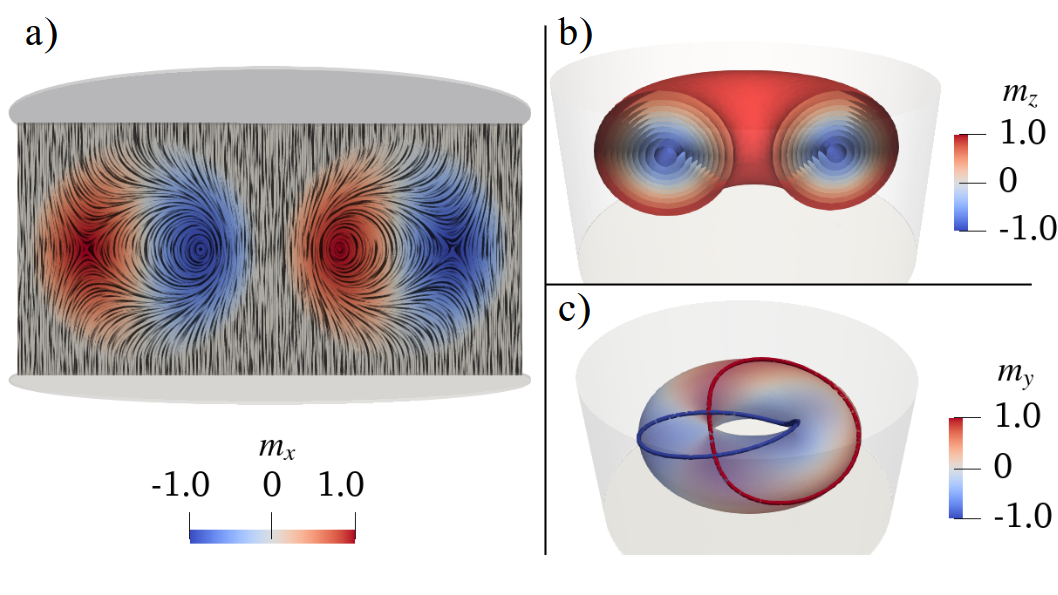}
\caption{Magnetic texture defined by Eqs.~(\ref{bimeron})-(\ref{bimeron_end}) in a cylinder of height $h=100$ and radius $r=100$, using $\alpha=\pi/2$, $m_0 = -1$, $u=1$, $v=1$, $r_\text{maj} = 50$ and $r_\text{min} = 45$.
(a) Magnetization distribution on a cutplane at $x=0$. In this streamline representation, the circular structures are magnetic vortices and the intersecting regions antivortices. The color code refers to the $x$ component, i.e., to the magnetization perpendicular to the cutplane.
(b) Perspective view on the toroidal-shaped nested set of $m_z$ isosurfaces, ranging from $m_z = -0.9$ to $m_z = 0.9$ by steps of $0.2$, sectioned along the $xz$-plane to reveal the internal structure. (c) Two intertwined preimages, $m_y = 1$ (red) and $m_y = -1$ (blue) which both belong to the $m_z = 0$ isosurface.}
\label{Hopfion_analytic}
\end{figure}

The characteristic toroidal $m_z$ isosurfaces composed of linked preimages displayed in Fig.~\ref{Hopfion_analytic} visually confirm that this analytic ansatz produces a magnetic texture with the defining features of a Hopfion.

\section{Numerical Results}
\begin{figure}[ht]
    \includegraphics[width=1\linewidth]{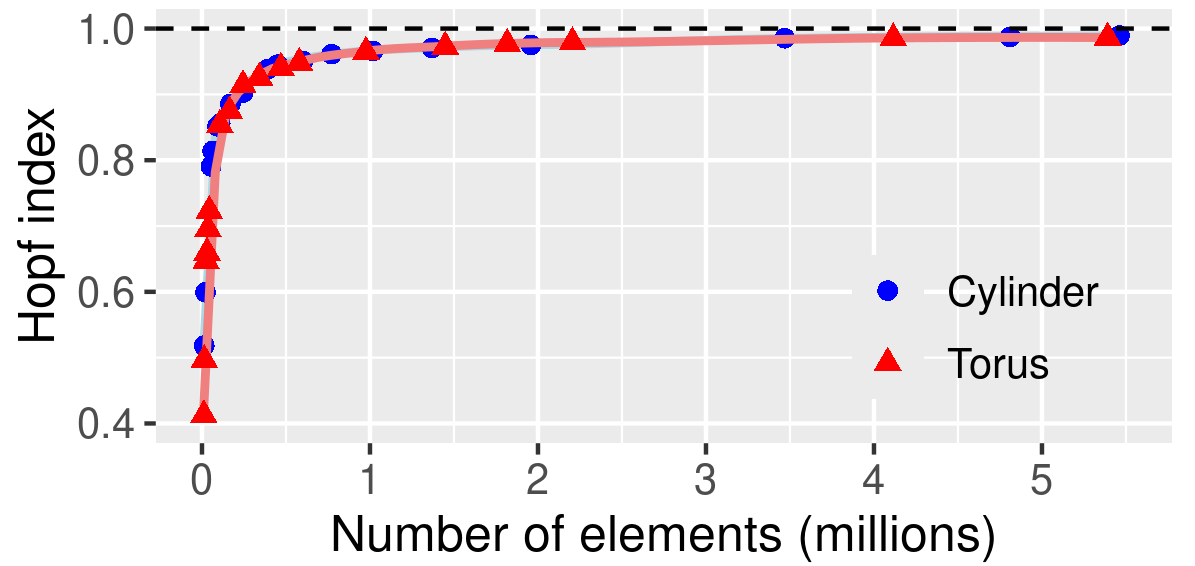}
    \caption{Calculated Hopf index for a cylinder and a torus geometry (see main text for details) as a function of the number of finite elements. In both cases, the expected value (dashed horizontal line) is reached asymptotically with increasing levels of discretization.}
    \label{hopf_vs_n}
\end{figure}

To test its correctness and accuracy, we apply the method described in section \ref{method_sect} and determine numerically the Hopf index of the analytically defined texture. With this model system, we expect to obtain a value close to unity in accordance with the topological characteristics of the Hopfion. Using Eqs.~(\ref{bimeron})-(\ref{bimeron_end}), we imprint the magnetic texture in a cylindrical geometry as illustrated in Fig.~\ref{Hopfion_analytic}.
The computed Hopf index is presented in Fig.~\ref{hopf_vs_n} as a function of the number of finite elements used in the calculation.

The ideal value $N=1.0$ is approached asymptotically as the number of elements increases. We observe this type of mesh-dependent convergence consistently across various shapes and geometries. For comparison, Fig.~\ref{hopf_vs_n} also displays data for the analytic Hopfion defined in a torus geometry with $r_\text{maj}=150$ and $r_\text{min}=60$, 
demonstrating equivalent results for different sample shapes and confirming the method’s ability to handle arbitrarily shaped 3D geometries. In the remainder of this section, we focus exclusively on data for the cylinder geometry. While these results confirm the method's overall correctness, they also highlight non-negligible numerical errors at low discretization levels and a sensitive dependence on the discretization density. 
 A similar behavior was reported by Liu {\em et al.} \cite{liu_binding_2018}, despite using an entirely different numerical scheme.
 These numerical inaccuracies can be problematic in the case of strong inhomogeneities, where they may preclude the precise quantitative identification of a texture's topological characteristics if the calculated Hopf index deviates too far from an integer value.

A primary source for the discretization errors is the numerical calculation of the topological current $\bm{J}$ as defined in Eq.~(\ref{topo_curr}) since it involves several products of various spatial derivatives. Further numerical errors accumulate as $\bm{J}$ is differentiated numerically on the right-hand side of the Poisson equation (\ref{Poisson_A}). In the context of the numerical analysis of the topological properties of Skyrmions, Kim {\em et al.} \cite{kim_quantifying_2020} introduced an elegant and efficient method to numerically determine the topological current with significantly higher accuracy by employing a lattice-based approach. In a recent article, Knapman {\em et al.}\cite{knapman_numerical_2025} demonstrated that this lattice-based method results in highly accurate numerical results for the case of Hopfions, too. The lattice-based approach, however, requires a coplanar distribution of discretization points---a criterion fulfilled in the case of finite-difference methods but inapplicable to finite-element methods, as discussed here, which use irregular discretization cells. 

\begin{figure}[ht]
    \includegraphics[width=1\linewidth]{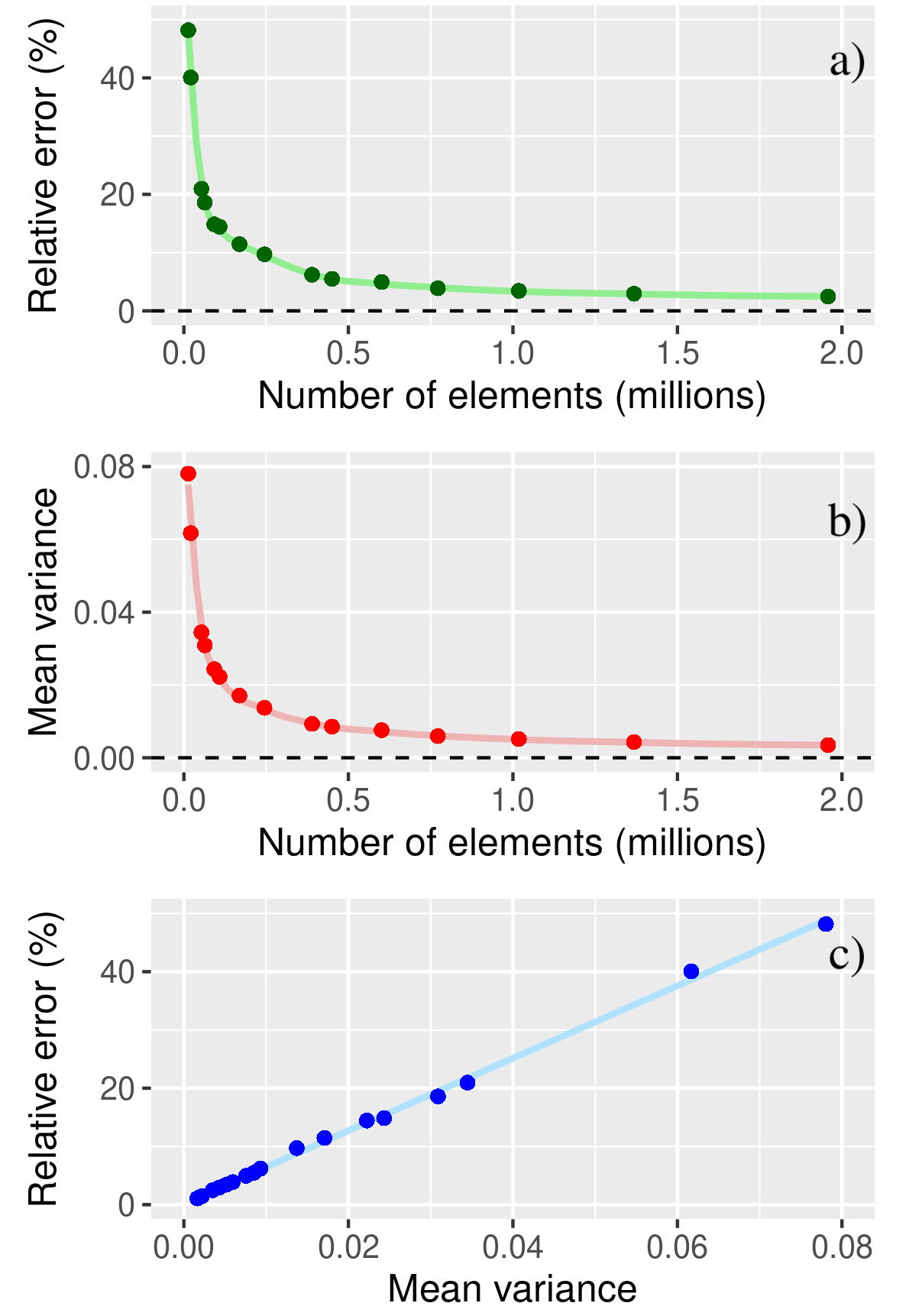}
    \caption{Discretization dependence of the relative error of the computed Hopf index (a) and average variance of the magnetization per finite element (b). The average variance and the relative error exhibit a nearly linear dependence (c).}
    \label{error_and_variance}
\end{figure}

To improve the numerical accuracy of our FEM approach, we employ a phenomenological {\em a posteriori} error correction scheme. This scheme is based on two observations we made consistently by studying a wide variety of situations. Firstly, the effect of the error is to {\em systematically underestimate} the absolute value of the Hopf index. Such behavior is easier to correct than stochastic fluctuations of the computed values around the exact ones. Secondly, we observe a clear correlation between the size of the error and the variance of the magnetization, as shown in Fig.~\ref{error_and_variance}.

Here, the variance is defined element-wise and provides a measure of the misalignment of neighboring magnetic moments. For each tetrahedral element $n$ with its four nodes $i=1\dots 4$, we can define the variance through

\begin{align}    
\sigma^2_n&=  \sum_{j=x,y,z} \langle m_j^2\rangle-\langle m_j\rangle^2 \\
& = \sum_{j=x,y,z}\left[\sum_{i=1}^{4} \frac{m_{i,j}^2}{4}-\left(\sum_{i=1}^4\frac{m_{i,j}}{4}\right)^2\right]\quad, \\
& = 1 - \sum_{j=x,y,z}\left[\left(\sum_{i=1}^4\frac{m_{i,j}}{4}\right)^2\right]\quad,
\end{align}

where the brackets $\langle \rangle$ denote the mean of the values at the nodes of element $n$. The value of $\sigma_n^2$ ranges from zero in the case of a homogeneous state to one in the case of pairwise antiparallel alignment of the four magnetic moments. More generally, the value of the variance correlates with the angle enclosed between neighboring magnetic moments within a finite element---a quantity known to be tied to discretization errors in numerical micromagnetics \cite{donahue_variational_1998}. It is therefore plausible to identify the variance as a measure or indicator for the occurrence of discretization errors. As shown in Fig.~\ref{error_and_variance}c), the relative error and the mean variance obtained for different meshes display a nearly linear dependence.

Based on these observations, i.e., the variance-dependent underestimation of the computed value, we design an error correction scheme to improve the calculated Hopf index $N_\text{calc}$. This is achieved through a function $f$ that yields a corrected value $N_\text{corr}=f(N_\text{calc}, \sigma^2)$. However, the mean variance depends on the number of elements used in the simulation, including those in regions without topological texture, making it unsuitable for a generally applicable correction algorithm. Instead, we use the local variance $(\sigma^2)_i$, ascribed to each node $i$ through a volume-weighted average of the variance in the adjacent elements, to locate and quantify numerical errors. Our correction method augments the absolute value of the calculated Hopf density $\varrho_i=(\bm{J}\cdot\bm{A})_i$ at each node by a factor that increases monotonically with the local variance. Despite the linear relationship between total error and mean variance illustrated in Fig.~\ref{error_and_variance}c), we did not achieve satisfactory results with a linear correction scheme based on the local variance. Rather than applying higher-order polynomial corrections, we improve the approximation by adopting a different functional form, where the Hopf density is increased at each node $i$ according to
\be \label{correction_eq}
\varrho_i\rightarrow\varrho_i\cdot\exp(\alpha\sigma^2_i)
\ee
with $\alpha = \text{const.}$ With this phenomenological approach, we obtain excellent results in a broad spectrum of situations by setting the constant's value to $\alpha=3.2$.

\begin{figure}[ht]
    \includegraphics[width=1\linewidth]{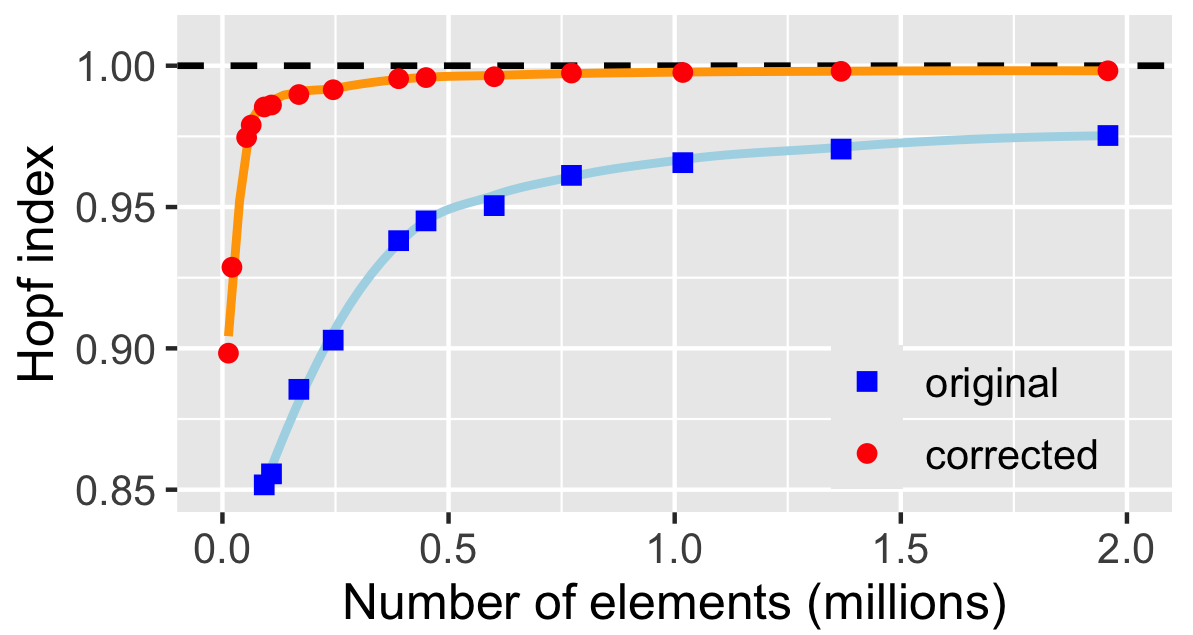}
    \caption{Impact of variance correction on the computed Hopf index and its discretization dependence.}
    \label{corrected_data}
\end{figure}

Fig.~\ref{corrected_data} illustrates the impact of the correction scheme on the discretization-dependent values of the calculated Hopf index and the relative error. Near-perfect agreement with the expected analytic value is obtained already at moderate discretization levels, and we rapidly obtain a substantial reduction of the relative error. For instance, already at a moderate number of about \num{390000} finite elements, the corrected Hopf index is $N_\text{corr}=0.995$, very close to the expected value of $N=1$, while the uncorrected index for the same mesh is $N_\text{orig}=0.938$. In this example, the correction decreases the relative error from $6.19\%$ to $0.46\%$.

Despite the remarkable accuracy obtained with this method, we emphasize that our phenomenological {\em a posteriori} correction is not suitable to sizably reduce the local error in the calculation of the gradients or the topological current. This is because the local discretization error, unlike the Hopf index, does not increase monotonically with the variance. It is subject to fluctuations in both magnitude and sign that cannot be efficiently reduced with the same approach. However, the {\em cumulative} discretization errors affecting the integral (or, in numerical terms, the sum over the elements) in Eq.~(\ref{hopf}) can be significantly reduced in this manner. 

We tested the method in several scenarios and geometries, considering different functional forms of the correction term and carefully tuning the constants involved. In all cases, the correction as described in Eq.~(\ref{correction_eq}) with $\alpha=3.2$ yielded excellent results, leading us to the conclusion that the choice of the exponential correction function and the value of the constant $\alpha$ are suitable for general cases, and are not specific to the texture discussed here. Eventually, we obtained enough confidence in this {\em a posteriori} correction to integrate it as a default operation into our algorithm to calculate the Hopf index. Hereafter, all reported Hopf index values are those corrected according to Eq.~(\ref{correction_eq}).

\section{Higher-order Hopfions\label{ho_hopf_sec}}
\begin{figure}[ht]
    \includegraphics[width=1.\linewidth]{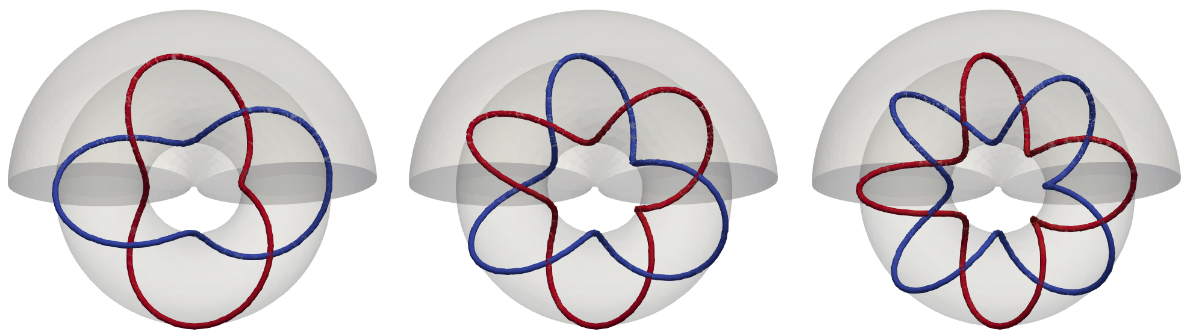}
    \caption{Preimage representation of higher-order Hopfions, defined through Eqs.~(\ref{bimeron})-(\ref{bimeron_end}) by setting $v=2, 3,$ and 4, in a torus geometry with $r_\text{maj}=150$ and $r_\text{min}=145$. The torus is discretized into approximately \num{99000} tetrahedral elements. The red and blue preimages refer to $m_y=1$ and $m_y=-1$ isolines, respectively, on the $m_z=0$ isosurfaces, represented by the semitransparent inner torus. The outer torus describing the geometric shape is cut at $x=0$ for better visibility.}
    \label{ho_hopf}
\end{figure}

The analytic formulation of Eqs.~(\ref{bimeron})-(\ref{bimeron_end}) enables the definition of higher-order Hopfion textures, {\em e.i.}, with a Hopf index larger than one \cite{rybakov_magnetic_2022}, by selecting appropriate values for the parameters $u$ and $v$. The parameter $u$ controls the number of bimerons in meridional cross-sections, as shown in Fig.~\ref{Hopfion_analytic}a) for $u=1$, while $v$ introduces a poloidal twist of the bimerons around the torus' core circle (the spine) at $\rho=0$.
Figure~(\ref{ho_hopf}) shows preimage representations of second-, third-, and fourth-order Hopfions in a torus geometry, obtained by setting $v=2,3,$ and 4, respectively (with $u=1, m_0 = -1$, and $\alpha=\pi/2)$. 

The poloidal twist generated by the parameter $v$ increases the vector field's degree of topological knottedness, as evidenced by the increasingly complex linking of preimage lines. Accordingly, our method yields the Hopf indices 1.995, 2.998, and 3.999 for these structures, respectively, matching these textures' ideal values 2, 3, and 4 very closely. These results further confirm our method's ability to reliably and quantitatively characterize the topological properties of complex magnetization vector fields in arbitrary geometries.

\section{Micromagnetic simulation}
Having established the method's reliability and accuracy using an idealized, analytically defined Hopfion texture, we now demonstrate an application example in a full-scale micromagnetic simulation. We consider a Hopfion bound in a cylindrical FeGe sample with height $h=\SI{100}{\nano\meter}$ and radius $r=\SI{100}{\nano\meter}$. The material parameters are $A=\SI{8.78}{\joule\per\meter}$, $M_s=\SI{384}{\kilo\ampere\per\meter}$, and $D=\SI{1.58}{\milli\joule\per\meter\squared}$, where $A$ is the ferromagnetic exchange constant, $M_s=\left|\bm{M}\right|$ is the saturation magnetization, and $D$ is the bulk-type Dzyaloshinsky-Moria constant. 
To stabilize the Hopfion at zero field\cite{liu_binding_2018}, we apply a Néel-type surface anisotropy $(K_{s} = \SI{10}{\milli\joule\per\meter\squared})$ on the top and bottom surfaces of the cylinder, which pins the magnetization along the $z$ axis at the caps.
We perform simulations using our custom-developed GPU-accelerated micromagnetic finite-element software \texttt{tetmag} \cite{hertel_tetmag_2023}.
To demonstrate the capacity of our method to reveal topological transitions, 
we investigate the quasistatic destruction of a Hopfion under an external magnetic field of increasing strength, monitoring field-induced changes in vector field topology through our method of calculating the Hopf index. 

The simulation is initialized with a Hopfion inside the cylinder, defined by the ansatz of Eqs.~(\ref{bimeron})-(\ref{bimeron_end}) with $\alpha=\pi/2$, $m_{0} = -1$, $u=1$, $r_\text{maj} = \SI{50}{\nano\meter}$ and $r_\text{min} = \SI{45}{\nano\meter}$, embedded into a homogeneous background magnetization $m_\text{z} = - m_0$. The numerical model employs a mesh with about 2 million irregularly shaped tetrahedral elements.

\begin{figure}[ht]    
    \includegraphics[width=\linewidth]{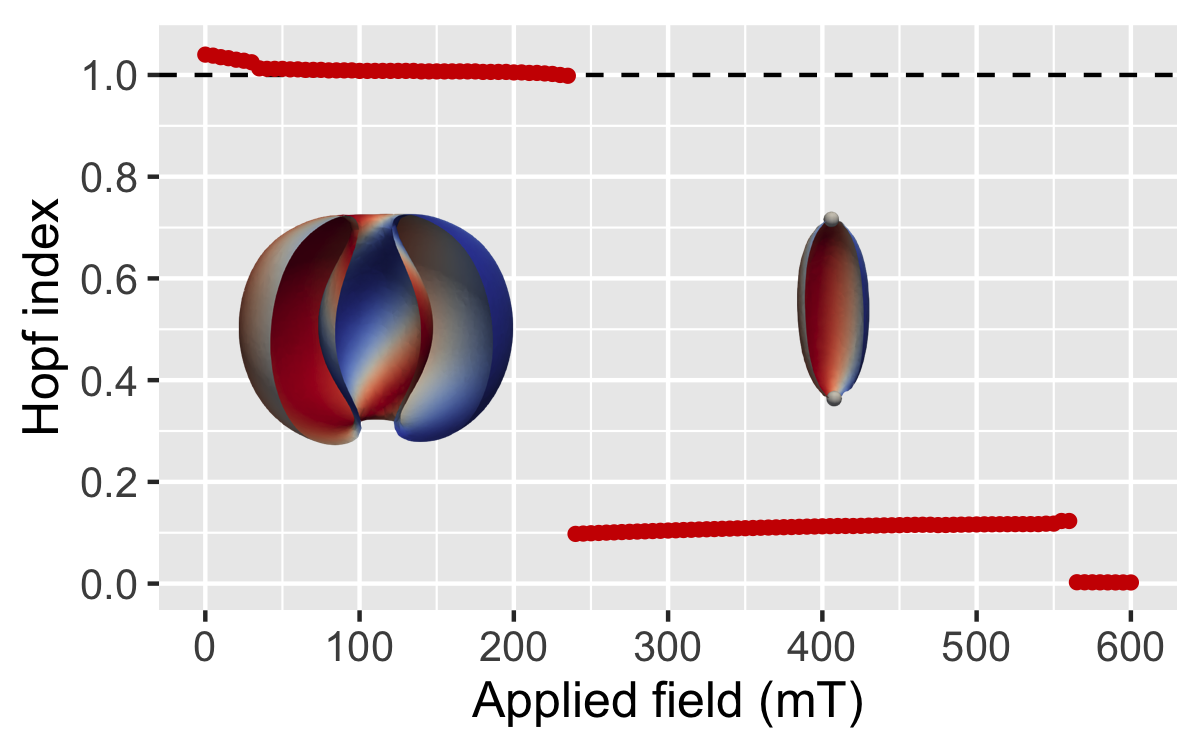}
    \caption{Calculated Hopf index as the magnetic structure changes under increasing field strength. Sudden changes of the Hopf index at specific field strengths indicate topological transitions. The insets are representative examples of the $m_z=0$ isosurfaces in the Hopfion and the toron configuration. The thin dashed line at 1.0 represents the expected ideal value for a Hopfion.}
    \label{fig: hopf_vs_h}
\end{figure}

We perform quasi-static relaxation simulations with an external field applied along the positive $z$-direction, varying from zero to \SI{600}{\milli\tesla} in steps of \SI{5}{\milli\tesla}. For each converged state, we calculate the Hopf index. The results are displayed in Fig.~\ref{fig: hopf_vs_h}, where three different regions are apparent, which are separated by two discontinuous changes in the Hopf index---the first at \SI{240}{\milli\tesla} and the second at \SI{560}{\milli\tesla}.  

\begin{figure}[ht]
    \includegraphics[width=1\linewidth]{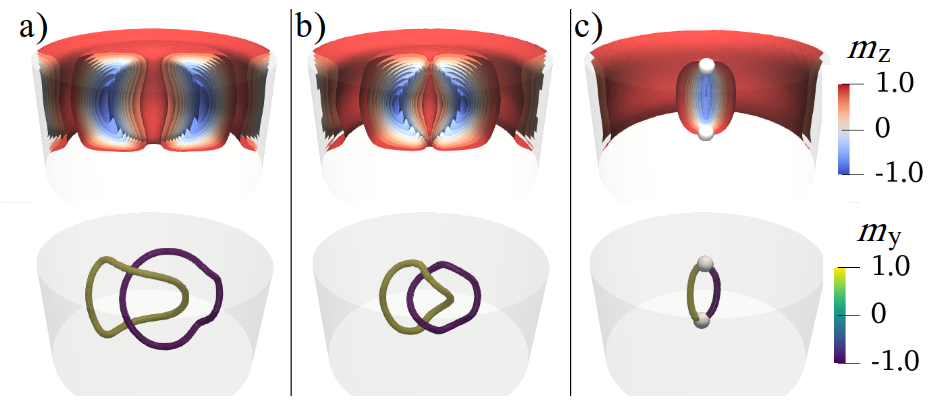}
    \caption{
    Isosurface (top row) and preimage (bottom row) representation of the magnetic structure in the FeGe cylinder at different field strengths. The cylinder is cut at the $x=0$ plane to display the inner structure. The Hopfion texture of panel (a), taken at $\mu_0H_\text{ext}=\SI{35}{\milli\tesla}$ and yielding a Hopf index $N=1.012$, shows the typical toroidal shape of the $m_z$ isosurfaces and linked, intertwined preimages in the isoline representation. b) Shortly before the destruction of the Hopfion, at $\mu_0H_\text{z}$ = \SI{235}{\milli\tesla}, the index slightly drops to $N=0.998$ while the shapes of the isosurfaces and the preimages are preserved, albeit more compressed compared to the low-field state. c) At $\mu_0H_\text{z}$ = \SI{560}{\milli\tesla}, the Hopfion has transformed into a toron, with Bloch points (represented as small spheres) on the top and bottom of the structure. In this configuration, the topological index evaluates to $N=0.12$.\label{fig:three graphs}}   
\end{figure}

Between \SI{0}{\milli\tesla} and \SI{235}{\milli\tesla}, the Hopf index remains close to 1, consistent with the expected value given the presence of a Hopfion.
At very low fields, the Hopf index exceeds this ideal value by up to about $4 \%$ due to spurious effects related to weakly developed textures near the cylinder's outer boundary, which are marginally connected to the Hopfion. 
The peripheral magnetization aligns more strongly with the applied field as the field increases beyond about \SI{35}{\milli\tesla} (see Fig~\ref{fig:three graphs}, a), causing these lateral textures to detach from the Hopfion, which becomes fully embedded into a homogeneous background magnetization $m_z = -m_0$. From \SI{35}{\milli\tesla} to \SI{235}{\milli\tesla}, the Hopf index stays nearly constant, with deviations below $\SI{1}{\percent}$. 
Despite significant field-induced changes, in particular a reduction in the Hopfion’s size (Fig.~\ref{fig:three graphs}), our correction method prevents any decrease in the Hopf index, even as local magnetic inhomogeneity increases. In this field range, where changes of the magnetization are reversible, the consistent Hopf index accurately reflects the topological equivalence of the textures, regardless of their field-induced deformation.

In the field range between \SI{240}{\milli\tesla} and \SI{560}{\milli\tesla}, the Hopf index remains almost constant at approximately 0.1; a significant decrease from its value in the first region. This abrupt reduction of the Hopf index at \SI{240}{\milli\tesla} marks the irreversible destruction of the Hopfion, which transforms into a toron as Bloch points form on the top and bottom end of an encapsulated core region in which the magnetization is oriented along the negative $z$ axis, oppositely oriented to the surrounding background (Fig~\ref{fig:three graphs}c). In the toron configuration, sometimes referred to as a coupled monopole-antimonopole pair\cite{liu_binding_2018}, the $m_z$ isosurfaces become homeomorphic to a sphere rather than a torus, and the previously closed, intertwined isolines in the preimage representation are now connected by two Bloch points. 
The change in vector field topology between the Hopfion and the toron configuration represents a topological transition that is characterized by an abrupt change in the Hopf index from 1 to 0.1. 
Like the first region, the Hopf index calculated in the toron regime remains constant over a wide range of field strengths, despite considerable field-induced modifications of the toron structure.

We emphasize that, while the change in the Hopf index indicates a topological change, the low value of 0.1 of the Hopf index does not allow definitive conclusions about the resulting magnetic structure beyond confirming the absence of a Hopfion.
 As discussed in section \ref{hopf_ind_sec}, the Hopf index calculation relies on a smooth magnetization vector field and a divergence-free topological current. These conditions are not met in the vicinity of Bloch points, which represent sources and sinks of the topological current\cite{liu_binding_2018}, invalidating the fundamental assumption $\bm{\nabla}\cdot\bm{J}=0$ required to introduce the vector potential $\bm{A}$ according to Eq.~(\ref{curlA}). Consequently, the Helmholtz decomposition of the topological current becomes $\bm{J}=\bm{\nabla}\times\bm{A}+\bm{\nabla}\Psi$, where $\Psi$ is a scalar potential accounting for the irrotational components. The vector potential $\bm{A}$ captures only the solenoidal part of $\bm{J}$, which is insufficient for a full description of the Bloch point topology. The stable Hopf index of approximately 0.1 in the toron regime across a wide field range suggests that the divergence-free part of $\bm{J}$ remains constant, potentially indicating a characteristic toron property. However, given the particular situation of a discontinuous magnetization vector field, we cannot rule out that spurious effects arising from the Bloch points may also contribute to the nonzero topological index. In any case, the fractional Hopf index calculated in this case cannot be interpreted topologically as a linking number. 

In the third region, from \SI{565}{\milli\tesla} to \SI{600}{\milli\tesla}, except for a slight rotation on the lateral surface, the magnetization is overall homogeneous along the field direction, resulting in a vanishing Hopf index.

\section{Conclusion}
We have developed a robust and accurate method for calculating the Hopf index in finite-element micromagnetic simulations. By employing the Biot-Savart form for the vector potential, our approach ensures gauge-invariant results even in multiply connected geometries. Applying our variance-based correction scheme to an analytically defined ideal Hopfion structure, we demonstrated its ability to significantly reduce numerical errors, which could otherwise be substantial in strongly inhomogeneous textures or exhibit slow mesh-size dependent convergence.

Through a micromagnetic simulation example involving a Hopfion's field-induced destruction, we confirmed the method's capacity to detect Hopfions and their absence, revealing topological transitions through abrupt changes in the Hopf index. Combined with the finite-element method's flexibility in modeling magnetic textures in arbitrarily shaped 3D nanostructures, this approach enables precise identification and quantification of topological features in complex 3D magnetic configurations. Our method provides a powerful tool for advancing the study of topological magnetism in micromagnetic simulations. While we have only demonstrated its effectiveness in micromagnetism, we are confident that the method can also be applied successfully in other scientific domains, particularly in other ferroic material systems.
\section*{Acknowledgements}
This work was supported by the French National Research Agency (ANR) under Contract No. ANR-22-CE92-0032 through the TOROID project co-funded by the Deutsche Forschungsgemeinschaft (DFG, German Research Foundation) Project No. 505561633 and by the France 2030 government investment plan managed by the French National Research Agency ANR under grant reference PEPR SPIN – [SPINTHEORY] ANR-22-EXSP-0009.

\end{document}